\documentclass[prl,aps,twocolumn,floatfix]{revtex4}
\usepackage{graphicx}
\usepackage{epsfig}
\def \ds {\displaystyle}
\def \hm {{\bf m}}
\def \D1 {\Delta_1 }
\def \ceff {c_{\rm eff}}

\def \vS {{\bf S}}

\def \Jc {J_c}

\def \hk {\hat {\bf k}}

\def \vj {{\bf j}}

\def \Emin {E^{\rm min}}
\def \kperp {{\bf k}_{\perp }}
\def \skperp {k_{\perp }}

\begin{document}

\date{\today}
\title{Nature of the Perpendicular-to-Parallel Spin Reorientation in a Mn-doped GaAs Quantum Well:  Canting or Phase Separation?}
\author{Randy S. Fishman$^1$, Fernando A. Reboredo$^1$, Alex Brandt$^{1,2}$, and Juana Moreno$^3$}
\affiliation{$^1$Materials Science and Technology Division, Oak Ridge National Laboratory, 
Oak Ridge, Tennessee 37831-6032}
\affiliation{$^2$Department of Physics and Astronomy, Minnesota State University Moorhead, Moorhead, MN 56563} 
\affiliation{$^3$Physics Department, University of North Dakota, Grand Forks, 
North Dakota 58202-7129}

\begin{abstract}

It is well known that the magnetic anisotropy in a compressively-strained Mn-doped GaAs film 
changes from perpendicular to parallel with increasing hole concentration $p$.  We study this 
reorientation transition at $T=0$ in a quantum well with Mn impurities confined to 
a single plane.   With increasing $p$, the angle $\theta $ that minimizes the energy $E$ 
increases continuously from 0 (perpendicular anisotropy) to $\pi /2$ (parallel anisotropy) within 
some range of $p$.  The shape of $\Emin (p)$ suggests that the quantum well becomes phase
separated with regions containing low hole concentrations and perpendicular moments 
interspersed with other regions containing high hole concentrations and parallel moments.  
However, due to the Coulomb energy cost associated with phase separation, the 
true magnetic state in the transition region is canted with $0 < \theta < \pi /2$.  
  
\end{abstract}
\maketitle

\newpage

During the past decade, the ferromagnetic transition temperature of Mn-doped GaAs 
films has been quickly approaching room temperature \cite{jung05}.   Theoretical studies 
of Mn-doped GaAs \cite{jung06} based on the Kohn-Luttinger (KL) model 
have been quite successful at modeling and predicting much of the behavior 
found experimentally.  In agreement with theory \cite{abol01,dietl01}, experiments show 
that at low temperatures, the magnetic anisotropy of films under compressive strain is 
perpendicular or out of the plane when the hole concentration $p$ is small, transforming to
parallel or in the plane as $p$ increases  \cite{saw04}.  Although Mn-doped GaAs 
quantum wells have also been studied theoretically for several years \cite{brey00,mac00}, 
little is known about the nature of the spin reorientation in a quantum well.  We show that
the spin reorientation in a quantum well happens in three stages:  for small hole concentrations, 
the angle $\theta $ of the magnetization with respect to the film normal is 0 so that the magnetization 
lies perpendicular to the plane;  for high hole concentrations, $\theta = \pi /2$ so 
that the magnetization lies in the plane of the quantum well.  In between, the 
moments are either canted with $0 < \theta < \pi /2$ or phase-separated with regions containing 
low hole concentrations and perpendicular moments interspersed with regions containing high
hole concentrations and parallel moments.

The magnetic anisotropy of a quantum well sensitively depends on the spin-orbit 
coupling.  In pure GaAs, the spin-orbit coupling plays two roles.  First, it
lowers the energy of the $j=1/2$ band compared to the $j=3/2$ band at the
$\Gamma $ point by about 320 meV.   Since the lower $j=1/2$ band is rarely occupied
by any holes, it is commonly ignored.  Second, spin-orbit coupling changes the energies 
of the $j=3/2$ holes so that heavy ($m_h=0.5m$) and light ($m_l=0.07m$) holes carry angular 
momentum $\vj \cdot \hk =\pm 3/2$ and $\pm 1/2$, respectively, along their momentum
direction $\hk $.   In the absence of elastic strain, the energies of the light and heavy hole bands 
in bulk GaAs are degenerate at the $\Gamma $ point.  

In a quantum well bounded by $z=\pm L/2$, the square of the $z$ 
component of the momentum is quantized.  For the two lowest wavefunctions
$\psi_1(z)=\sqrt{2/L}\cos(\pi z/L)$ and $\psi_2(z)=\sqrt{2/L} \sin (2\pi z/L)$ of the 
quantum well, $\langle n| k_z^2| m \rangle =(n\pi /L)^2\delta_{nm}$. 
Due to the difference between the light and heavy hole masses, the confinement 
of the holes in a quantum well breaks the degeneracy of the $j=3/2$ bands at the $\Gamma$ point
with $\kperp =0$.   To simplify the following discussion, we shall discuss 
hole rather than electron bands so that hole energies increase quadratically 
like $\skperp^2$ for small $\skperp $.  Including the effects of lattice strain, the energy gap between 
the $j_z=\pm 3/2$ and $\pm 1/2$ sub-bands of $\psi_n(z)$ at the $\Gamma$ point is
\begin{equation}
\label{Delc}
\Delta_n = \frac{\langle n|k_z^2|n \rangle }{2}\biggl\{ \frac{1}{m_l}
-\frac{1}{m_h}\biggr\} -2b_d Q_{\epsilon},
\end{equation}
where $Q_{\epsilon }= \epsilon_{zz}-(\epsilon_{xx}+\epsilon_{yy})/2$ and $b_d \approx -1.6$ eV
is the deformation potential \cite{dietl01}.   Hence, compressive strain 
($Q_{\epsilon } > 0$) plays qualitatively the same role as carrier confinement
within the quantum well.  As the quantum well becomes narrower, the dominant
contribution to this splitting comes from the confinement of the holes and not from 
strain, which is typically less than 0.5\%.   For a quantum well with 20 layers or fewer,
the contribution of strain to $\Delta_n $ can be safely neglected.   On the other
hand, GaAs films with $\Lambda $ larger than about 50 cannot be treated as quantum wells 
because too many wavefunctions $\psi_n(z)$ would be required. 
In such films, the main contribution to the band splitting comes from 
strain \cite{abol01} rather than the confinement of the holes.
Regardless of their origin, the energy splittings $\Delta_n $ produce a gap in 
the spin-wave spectrum that allows a two-dimensional layer of Mn spins to order 
ferromagnetically \cite{melk07}.

The carriers in the lower, $j_z=\pm 3/2$ and upper, $j_z=\pm 1/2$
sub-bands have bandmasses $m_a=0.089m$ and $m_b=0.197m$, respectively,
where $4/m_a=3/m_l+1/m_h$ and $4/m_b=1/m_l+3/m_h$ \cite{dietl01,melk07}.    
The coupling of the $j=3/2$ holes with the Mn spins is included by treating the $S=5/2$ Mn spins 
classically and by assuming that Mn impurities with concentration $c$ 
are restricted to the $z=0$ plane.  We shall denote $p$ as the number of holes 
for each of the $N$ cations in the central plane.   For small $p$, the holes occupy a small portion 
of the $\kperp $ Brillouin zone centered around $\kperp =0$.  Since the holes then
interact with many different Mn moments, the precise locations and
structure of the Mn impurities are not important and their interactions with the 
holes may be treated within a mean-field approximation.  The exchange coupling of the 
Mn spins $\vS_i=S\hm_i$ with the holes is then given by $V=-2J_c\sum_i \hm_i \cdot \vj_i $, 
where $\vj_i$ are the hole spins and the sum is over all Mn sites.  
Comparing $V$ with the potential used in Refs.\cite{abol01} 
and \cite{dietl01}, we obtain the exchange coupling 
$J_c = (S/4j)\beta N_0  2c/\Lambda $, where $\beta N_0 \approx 1.2$ eV 
is estimated from photoemission measurements \cite{oka98} and $\Lambda =L/b$ is the 
number of layers in the quantum well.  Here, $b\approx 4 \AA $ is the Ga lattice constant in the
$z=0$ plane.   It follows that $J_c\approx 1 \, c/\Lambda $ eV.  

\begin{figure}
\includegraphics *[scale=0.35]{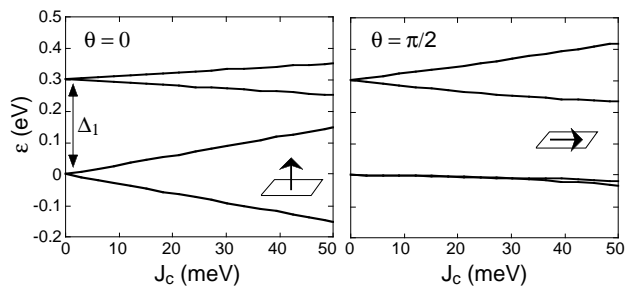}
\caption{
The energies of the bottom of the carrier bands with $\kperp = 0$ versus $\Jc $ when only
$\psi_1(z)$ is considered.  The chemical potential increases with the hole filling.
}
\end{figure}

While our results are supported by numerical calculations that include both wavefunctions 
$\psi_1(z)$ and $\psi_2(z)$, a qualitative understanding of the
magnetic anisoptropy can be obtained from a simplified model that considers only $\psi_1(z)$.
Due to the typically small Mn concentrations,
the demagnetization fields can be neglected compared to the anisotropy introduced by the
electronic band structure.  If the Mn moments are tilted an angle
$\theta $ away from the $z$ axis, then to linear order in $J_c$, the two lower sub-bands are 
split by $\pm 3J_c$ ($\theta =0$) or 0 ($\theta = \pi /2$), 
as shown in Fig.1 for $\Lambda =10$.  By contrast, the two upper bands are split by
$\pm J_c$ ($\theta =0$) or $\pm 2J_c$ ($\theta = \pi /2$).  When only the lowest sub-band is
populated by holes, the energy difference $(E(\theta=0)-E(\theta=\pi /2))/N$ is of order
$-J_c$, so that the anisotropy is perpendicular.   When the two lower sub-bands are 
both occupied by holes, the energy difference is of order $-J_c^2 mb^2/c$ because more
holes occupy the lowest sub-band and perpendicular anisotropy still dominates.
But when holes begin to occupy the lower of the two upper sub-bands, the energy 
difference becomes positive and of order $J_c$ due to the larger splitting of the 
upper sub-bands when $\theta =\pi /2$.  Assuming that $J_c \ll \D1 $, the reorientation
transition occurs close to the filling where the chemical potential 
$\mu \approx \pi p/m_a b^2$ crosses $\D1 \approx (\pi^2 /2\Lambda^2 b^2)(1/m_l-1/m_h)$.  
Very roughly, this implies that the transition from perpendicular to parallel anisotropy
occurs when $p \sim 1/\Lambda^2$, independent of the Mn concentration.

The scenario of magnetic reorientation would be reversed when tensile strain overcomes 
the effects of carrier confinement so that $\D1 < 0$.  Because perpendicular or 
parallel anisotropy dominates for compressive or tensile strain only at very low hole
concentrations, the behavior of the anisotropy at higher hole 
concentrations has led to the often-heard statement \cite{shen97, liu03, thev06} that 
compressive or tensile strain is associated with parallel or perpendicular anisotropy.

\begin{figure}
\includegraphics *[scale=0.5]{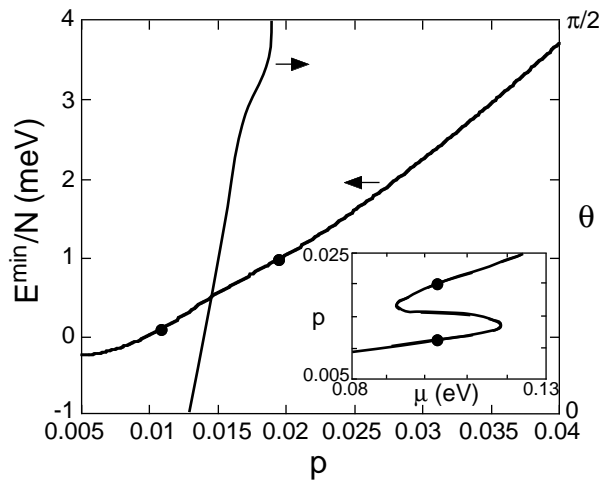}
\caption{
The minimum energy $\Emin /N $ and the angle $\theta $ versus $p $ for $c=0.4$
and $\Lambda = 10$.   Inset is a plot of $p $ versus $\mu $, revealing an "S" shaped
curve.  A Maxwell construction yields phase separation between the two solid circles. 
}
\end{figure}

We now examine more closely the details of the spin reorientation.   With both 
wavefunctions $\psi_1(z)$ and $\psi_2(z)$ included,  there are 8 hole bands rather 
than 4 for every $\kperp $ point.  
Since $\psi_2(0)=0$, the Mn spins only couple to the holes in $\psi_1(z)$,
with projection $\vert \psi_1(0)\vert^2 = 2/L $.  The wavefunctions $\psi_1 (z)$
and $\psi_2(z)$ are coupled by the off-diagonal terms in the KL
Hamiltonian with matrix elements proportional to $\langle n|(k_x\pm
ik_y)k_z|m\rangle $, which vanishes for $n=m$ but is given by
$-(8i/3L)(k_x\pm i k_y)$ for $n=1$ and $m=2$.
The energy $E(J_c, \theta )$ of the KL plus exchange (KLE) model is 
obtained by first diagonalizing the Hamiltonian written as an 8 by 8 matrix in
$j=3/2$ and $n,m=1,2$ space.   The resulting eigenvalues are integrated over 
$\kperp $ up to the Fermi level \cite{melk07}.

Fixing $c$ and $p$, the energy $E$ is then minimized with respect to $\theta $.  
For $c=0.4$, $\Emin /N $ is plotted versus $p$ in Fig.2.   As shown, 
the angle $\theta $ corresponding to the minimum energy smoothly increases 
from 0 at $p_1=0.0128$ to $\pi /2$ at $p_2=0.019$.   Of course, the chemical
potential $\mu $ satisfies the condition $d\Emin /dp = N\mu $.  We have plotted $p$ versus 
$\mu $ in the inset to Fig.2.   This ``S'' shaped curve is typical of phase separation,
where regions of different $\{\theta ,p  \}$ coexist.   The phase
with high or low hole density is metastable so long as $d^2\Emin /dp^2 > 0$ or
$dp /d\mu > 0$, while the phase is
unstable when $dp /d\mu < 0$.  Performing
a Maxwell construction for the data in Fig.2, we find that phase separation
occurs between regions with hole concentration $p_l = 0.0113 < p_1 $ and $\theta =0$
and regions with $p_u = 0.02 > p_2 $ and $\theta=\pi /2$.  The phase-separated
region is bracketed by the solid circles in Fig.2.   As the hole concentration 
increases from $p_l$ to $p_u$, the total area of the regions with parallel or
perpendicular anisotropy increases or decreases, respectively.  A phase-separated
phase is stable only if the KLE Hamiltonian is expanded in more than 
one $\psi_n(z)$.

\begin{figure}
\includegraphics *[scale=0.5]{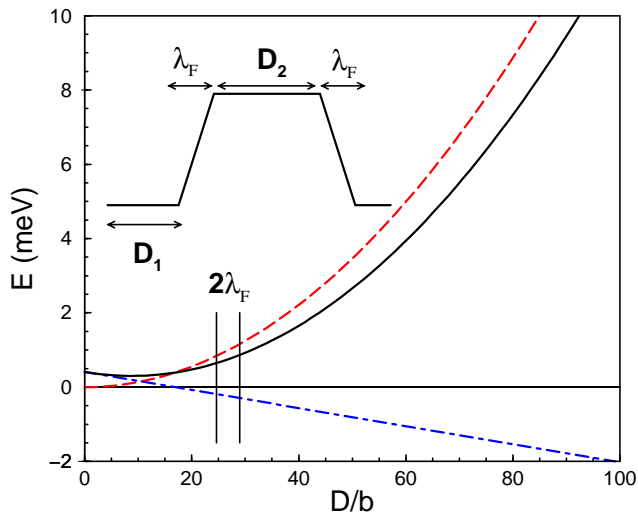}
\caption{(Color online) The total energy gain of a phase separated mixture 
(blue dot-dashed curve), the Coulomb cost associated with the domain wall 
(red dashed curve), and the total energy when both contributions are added 
(black solid curve) versus the size of the region for $c=0.5$ and $\Lambda =10$.  
The higher and lower values of $2\lambda_F$ are associated with perpendicular and parallel 
Mn moments, respectively.  The results below $2\lambda_F$ have no physical meaning since
$D_1$ and $D_2$ would have to be negative.}
\end{figure}

However, phase separation of the quantum well into regions with low and high
hole concentrations costs Coulomb energy, which was not taken into account by
the KLE model.   The size of the phase-separated regions will be determined
by a balance between the cost in Coulomb energy and the energy gained 
by phase separation.  The Coulomb and phase-separation energies are estimated 
by supposing that a charge-density wave in the Mn plane oscillates between fillings 
$p_l$ and $p_u$ within a rectangular region in the $z=0$ plane of length $D$ and width $b$.   
As described in the inset to Fig.3, a region of length $D_1$ and filling $p_l$ and a region of 
length $D_2$ and filling $p_u$ are separated by an interface of length
$\lambda_F$.  The total length of the rectangular region is $D=D_1+D_2+2\lambda_F$,
which is measured in units of the in-plane lattice constant $b$.   For overall filling $p$,
charge conservation requires that $pD=p_lD_1+p_uD_2+\lambda_F(p_l+p_u)$.

To calculate the energy gained by creating a phase-separated mixture, we subtract the 
energy of a uniform phase with $p=(p_l+p_u)/2$, yielding the blue dot-dashed curve in 
Fig.3 \cite{unphys}.  The dielectric constant for GaAs is used to evaluate the Coulomb 
energy cost, given by the red dashed curve in Fig.3.
The Fermi wavelength $\lambda_F=2 \pi/k_F$ is evaluated for this uniform filling $p$ 
with both Mn orientations.   While the perpendicular $k_F$ is a bit smaller than the 
parallel $k_F$, the calculated total energies are both positive.  We conclude that 
phase separation is prevented by the cost in Coulomb energy for $c=0.5$ and 
that the Mn moments will be canted for fillings between $p_1$ and $p_2$.  This is also 
the case for smaller Mn dopings which are slightly more unfavorable to phase separation.
In the homogeneous phase with filling $p$, the canting angle $\theta $ is not
affected by the Coulomb energy and $\theta (p)$ can be taken directly from
our results.  

Since the long-range electric field contribution to the Coulomb energy may be suppressed in a 
double quantum-well, where the excess electronic charge on one quantum well
is offset by the deficit on the other, a double quantum-well may exhibit phase separation 
rather than canting.  By tuning the distance between the individual quantum wells, 
it may be possible to control the transformation from canting to phase separation.  The 
dependence of the hole density on the magnetization angle will induce a coupling between 
spin- and charge-density excitations analogous to the ones observed in electronic 
systems \cite{giu04}.  The emergence of phase separation will soften the transverse 
charge-density mode in a double quantum well.

\begin{figure}
\includegraphics *[scale=0.42]{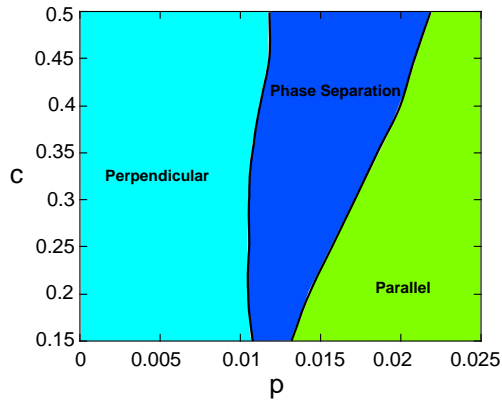}
\caption{
(color online) The phase diagram of the magnetic orientation in a quantum well with $\Lambda =10$, 
for Mn concentration $c$ versus holes per cation $p$, 
showing regions of perpendicular and parallel anisotropy, as well as a possible phase-separated
region.
}
\end{figure}

In Fig.4, we plot the phase diagram of the magnetic orientation in the quantum well,
leaving open the possibility of phase separation.  The canted region is just a bit
narrower than the phase-separated region shown in the figure.  The lower bound to the 
canted region is given fairly accurately by $p_1 \approx 0.012$ for all $c$.
By contrast, the upper bound $p_2 $ increases from 0.0135 at $c=0.15$
to 0.018 at $c=0.5$.    Hence, both the phase-separated and canted regions grow 
as the Mn concentration increases.  

Due to the large size of the gap energy $\Delta_1$, we 
do not expect our results to change very much with increasing temperature so long as
$T \ll \D1 \approx 29/\Lambda^2$ eV.   When the Mn concentration $c(z)$ is 
distributed along the width of the quantum well, our results with effective Mn concentration
$\ceff =(L/2)\int_{-L/2}^{L/2} dz \vert \psi_1(z)\vert^2 c(z)$ will be unchanged 
provided that $(L/2)\int_{-L/2}^{L/2}  dz\vert \psi_2(z)\vert^2 c(z)\ll \ceff $.  We have also
examined the consequences of moving the Mn plane from the center of the quantum 
well to $z'=\pm L/6$.  Since $\vert \psi_1(z')\vert^2=\vert \psi_2(z')\vert^2$, the 
Mn moments in this plane couple equally to the holes of both wavefunctions.   For $c=0.4$,
the canted region between $p_1=0.048$ and $p_2=0.064$ is shifted substantially upwards 
by the displacement of the Mn plane.

Although the spin reorientation transition has not been studied in a quantum well, 
several experiments have been performed on thin films.  By measuring the remanent 
magnetization along different field directions, Sawicki {\em et al.} \cite{saw04} estimated 
that the change from perpendicular to parallel magnetic anisotropy in 400 nm films with 
$c=0.03$ or 0.05 occurs when $p\approx 10^{20}$ cm$^{-3}$ or 0.0045 holes per cation.  
Using angle-dependent x-ray magnetic circular dichroism to study 50 nm films with $c=0.02$ 
or 0.08, Edmonds {\em et al.} \cite{edm06} observed this reorientation transition at a hole 
concentration of about 5 times higher or 0.0225 holes per cation.   

The hole concentration in a quantum well may be controlled using a field-effect 
transistor, such as the one built by Ohno {\em et al.} \cite{ohno00}.  
Of course, the canting or phase separation of the Mn moments would be easier to 
observe if the spin-reorientation transition happened at higher values of $p$.   As discussed
above, the easiest way to enhance the hole filling of the spin reorientation in a narrow
quantum well is to displace the Mn plane by 1/6 of the quantum-well width.  The relevant 
ingredients of the spin-reorientation transition are the single-particle gap $\Delta_1$ , 
the spin-orbit coupling in the valence band, and the magnetic coupling of the Mn ions with 
holes.  Accordingly, we believe that our predictions extend well beyond the simplest 
confinement potential discussed here to a broad class of quantum-well potentials.

This research was sponsored by the U.S. Department of Energy Division of 
Materials Science and Engineering under contract 
DE-AC05-00OR22725 with Oak Ridge National Laboratory, managed by UT-Battelle, LLC..
This research was also supported by NSF awards DMR-0453518, DMR-0548011, 
EPS-0132289 and EPS-0447679. Research carried out in part at the 
University of North Dakota Computational Research Center, supported by NSF 
EPS-0132289 and EPS-0447679.


\begin{references}
\bibitem{unphys}  The energy gained due to phase separation is
\begin{eqnarray}
&\Delta E_{{\rm PS}}(D)=\ds\frac{2\lambda_F}{p_u-p_l} \int_{p_l}^{p_u}dp'E(p')\nonumber \\ &+
E(p_l)D_1+E(p_u)D_2-E(p)D.
\end{eqnarray}
In the unphysical limit $D\rightarrow 0$, $D_1$ and $D_2 \rightarrow 
-\lambda_F$ and $\Delta E_{{\rm PS}}(0)>0$.  Although the 
intercept $\Delta E_{{\rm PS}}(0)$ is proportional to $\lambda_F$, it hardly
differs for the two possible values of $\lambda_F$ for parallel or perpendicular moments
so their average is used to plot the blue dot-dashed curve in Fig.3.
\end{references}
\end{document}